# Photometric Investigation of the Total-Eclipsing Contact Binary V12 in the Intermediate-age Open Cluster N G C 7789

Qian, S.-B.[1,2,3], Essam, A.[4], Wang, J.-J.[1,2,3], Ali, G. B.[4], Liu, L.,[1,2], and Haroon, A.-A.[4,5]

## A B S T R A C T

NGC7789 is an intermediate-age open cluster with an age that is similar to the mean age of contact binary stars. V12 is a bright UMa-type binary star with an orbital period of 0.397day. The first complete light curves of V12 in V, R, and I bands are presented and analyzed with the W-D (Wilson and Devinney) method. The results shows that V12 is an intermediate-contact binary ( $f$ = 43(±2.2)%) with a mass ratio of 3.848, and it is a W-type contact binary where the less massive component is slightly hotter than the more massive one. The asymmetry of the light curves are explained by the presence of a dark spot on the more massive component. The derived orbital inclination ( $i$ = 83.6°) indicates that it is a total eclipsing binary, which suggests that the determined parameters are reliable. The orbital period may show a long-term increase at a rate of $\dot{P}$ = +2.48(±0.17) × $10^{-6}$ days/year that reveals a rapid mass transfer from the less massive component to the more massive one. However, more observations are needed to confirm this conclusion. The presence of an intermediate-contact binary in an intermediate-age open cluster may suggest that some contact binaries have a very short pre-contact timescale. The presence of a third body or/and stellar collision may help to shorten the pre-contact evolution.

*Subject headings:* Stars: binaries : close – Stars: binaries : eclipsing – Stars: individuals (V12) – Stars: evolution

[1]Yunnan Observatories, Chinese Academy of Sciences (CAS), P.O. Box 110, 650011 Kunming, P.R. China (e-mail: qsb@ynao.ac.cn)

[2]Key laboratory of the structure and evolution celestial bodies, Chinese Academy of Sciences, P. O. Box 110, 650011 Kunming, P. R. China

[3]University of the Chinese Academy of Sciences, Yuquan Road 19#, Sijingshang Block, 100049 Beijing, P. R. China

[4]National Research Institute of Astronomy and Geophysics, Department of Astronomy, Helwan, Cairo, Egypt

[5]Astronomy Department, Faculty of Science, King Abdul Aziz University, Jeddah, KSA

# 1. Introduction

NGC 7789 is a rich and intermediate-age open cluster. Burbidge & Sandage (1958) determined the colors and magnitudes of nearly 700 stars in the cluster. A proper motion study made by McNamara & Solomon (1981) yielded 679 probable member stars of in NGC 7789. This cluster has been the target of several searches for variable stars. Jahn et al. (1995) made the first time-resolved CCD photometry which resulted in the discovery of 15 variables in the central part, among which most of the variables are of eclipsing type. Later, Kim, Kim & Park (1999) detected 16 variables, including eight suspected variables and three old objects previously discovered by Jahn et al. (1995). CCD photometric monitoring on NGC 7789 was carried out by Mochejska & Kaluzny (1999) who observed this cluster in two different fields (central part and an extended area). A total of 45 variables were found, 31 in the central part and 14 in the extended area. As the second target of the BATC variable searching program in open clusters (Zhang et al. 2002), NGC 7789 was monitored in 2000 based on time resolved CCD photometry. 28 new variable stars were discovered including 23 eclipsing binaries. Most of the eclipsing variables are of W UMa type with periods shorter than a day.

W UMa-type variable stars are interacting binaries where both components are over-filling their critical Roche lobe and share a common convective envelope. Photometric investigation of total-eclipsing contact binary stars in stellar clusters is very important to understand their formation and evolution. A few W UMa-type binaries (e.g., V22 and V31) in NGC 7789 have been studied (e.g., Sriram et al. 2010; Kiron et al. 2011). However, those targets are partial-eclipsing binaries and the original photometric data showed a large scatter. Therefore, photometric parameters can not be determined in a higher precision. V12(=2MASSJ23581634+5631202) is one of the bright eclipsing binary in NGC 7789 that was discovered by Zhang et al. (2002). It is a typical W UMa binary with an orbital period of 0.3917day. Recently, several contact binaries in different ages inside open clusters have been investigated (e.g., Qian et al. 2006, 2007; Liu et al. 2007, 2008, 2011; Zhu et al. 2014). The light curve of V12 derived Zhang et al. (2002) indicates that photometric parameters of the W UMa-type binary can be obtained in high precision because of the brightness and total eclipses. Therefore, it was included in our photometric-investigation program of total eclipsing binaries in stellar clusters.

# 2. New CCD photometric observations and orbital period changes

Photometric observations of the eclipsing binary system V12 in the intermediate-age open cluster NGC 7789 were obtained on $8^{th}$ and $9^{th}$ of October, 2013 by using EEV CCD 42-40 camera attached to the Newtonian focus of the 1.88-m Kottamia reflector telescope in Egypt (see Azzam et al. 2010). The CCD 42-40 camera has a format of 2048*2048 pixels with scale equals $0.\!\!^{''}305$/pixel that was cooled by liquid nitrogen to $-125C^\circ$. During the observations, the B, V, R, and I wide pass-band filters that are close to the standard Jonson system were used. PHOT (measure magnitudes for a list of stars) of the aperture photometry package of IRAF [1] was used to reduce the those CCD images. Two stars, GSC1-4009-02036 and TYC4009-939-1, were chosen as the comparison and the check stars. They are very close to V12 and their positions are shown in Fig. 1 where "V12" refers to the eclipsing binary, "C2" to the comparison star, and "C1" to the check star, respectively. The coordinates, magnitudes, and colors of these stars are listed in Table 1. The brightness and color of the comparison star are very similar to those of V12.

Table 1: Coordinates of V12, the comparison, and the check stars.

| Targets | Names | $\alpha_{2000}$ | $\delta_{2000}$ | $B$ | $B - V$ |
|---|---|---|---|---|---|
| V12 | 2MASSJ23581634+5631202 | $23^h58^m16.3^s$ | $56°31'20.3''$ | 13.873 | 0.636 |
| The comparison (C2) | GSC1-4009-02036 | $23^h58^m12.5^s$ | $56°32'17.8''$ | 13.874 | 0.596 |
| The check (C1) | TYC4009-939-1 | $23^h58^m20.2^s$ | $56°32'34.9''$ | 11.818 | 0.280 |

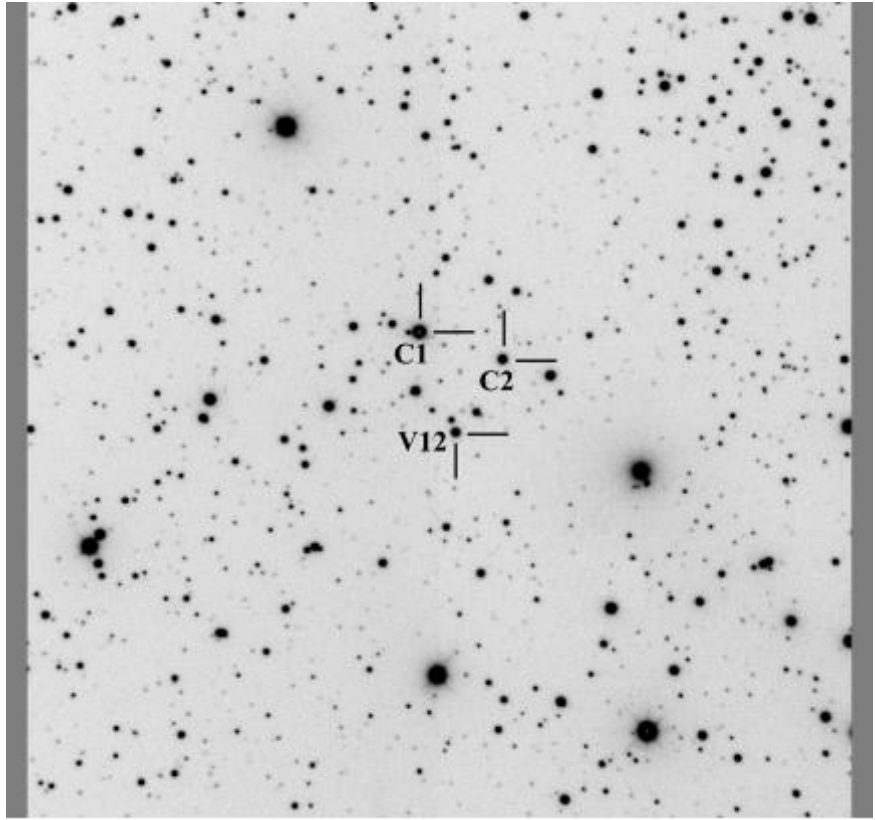

Fig. 1.— One of the CCD images of V12 in the intermediate-age open cluster NGC 7789 obtained by using the 1.88-m Kottamia reflector telescope in Egypt. C2 and C1 are the comparison and the check stars, respectively. North is up and East is to left.

The light curves in V, R, and I bands obtained on 8 and 9 October, 2013 are displayed in Figs. 2 and 3. As shown in the two figures, the light curves are typical EW-type where the brightness varies continuously and the depths of both light minima are nearly the same. The nearly flat eclipse bottom reveals that V12 is a total eclipsing binary. The amplitudes of the light variation are about $\approx$ 0.50mag in V, R, and I bands. The light curves show a positive O'Connell effect and the light maxima following the primary minima are higher than the other ones (O'Connell, 1951). Three epochs of eclipse times in different bands were determined with the Kwee & van Woerden (1956) method. The mean values are listed in Table 2.

The first linear ephemeris of V12 was determined by Zhang et al. (2002),

$$Min.I(HJD) = 2451812.146 + 0^d.3917 \times E, \qquad (1)$$

where E is the cycle number. HJD2451812.146 is the initial epoch, while $0^d.3917$ is the orbital period. Then one eclipse time, HJD2453765.5186, was determined by Biro et al. (2006). The $(O - C)$ values with respect to the linear ephemeris derived by Zhang et al.

---

[1]IRAF (an acronym for Image Reduction and Analysis Facility) is a collection of software written at the National Optical Astronomy Observatory (NOAO) geared towards the reduction of astronomical images in pixel array form.

Table 2: New CCD times of light minimum.

| HJD (days) | Errors (days) | Filters |
|---|---|---|
| 2456574.46377 | 0.00070 | BVRI |
| 2456575.44504 | 0.00030 | VRI |
| 2456575.24687 | 0.00059 | VRI |

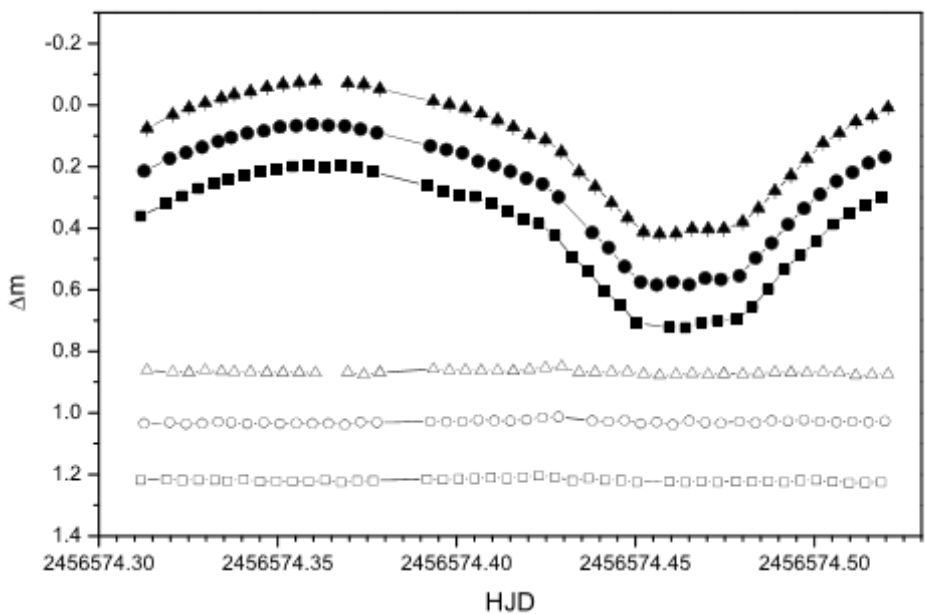

Fig. 2.— CCD photometric light curves of V12 observed on 8 October, 2013. Solid squares, circles, and triangles refer to V+0.1, R, I-0.1, respectively. Magnitude differences between C2 and C1 are also displayed in the figure as open squares, circles, and triangles for V, R and I band respectively.

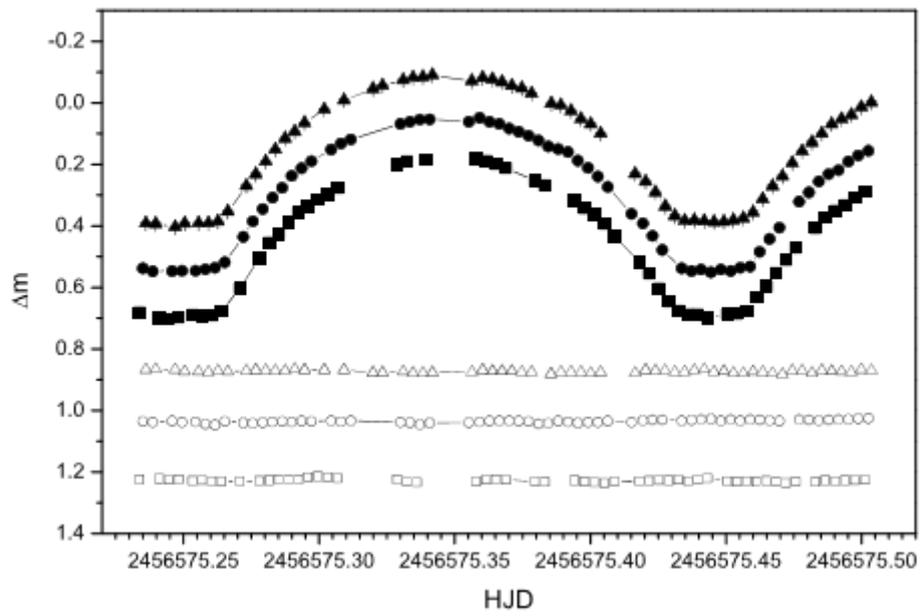

Fig. 3.— The same as those in Fig. 2 but observed on 9 October, 2013.

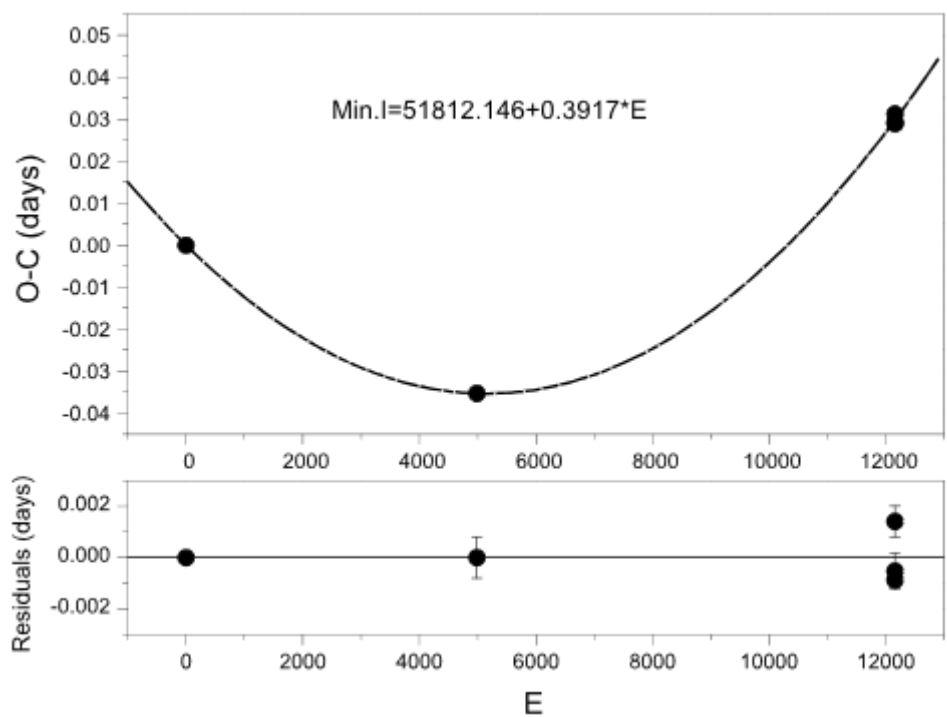

Fig. 4.— O-C diagram of V12 with respect to the linear ephemeris in Eq. (1). The upward parabolic variation (solid line) reveals a long-term increase in the orbital period. The residuals after the increase was subtracted are shown in the lower panel.

(2002) were computed and the corresponding ($O - C$) curve is displayed in Fig. 4 along with the epoch number E. As displayed in the upper panel of Fig. 4, the general trend of the O-C curve shows an upward parabolic change indicating an increase in the orbital period. By using a least-squares method, we yielded,

$$
\begin{aligned}
Min.I \quad = \quad & 2451812.1460(\pm 0.0004) \\
+ \quad & 0.3916863(\pm 0.0000001) \times E \\
+ \quad & 1.33(\pm 0.09) \times 10^{-9} \times E^2.
\end{aligned} \qquad (2)
$$

The quadratic term in Eq. (2) reveals a linear increase at a rate of $\dot{P} = +2.48(\pm 0.17) \times 10^{-6}$ days/year (or 21.4s in about 100years). The solid line in the upper panel of Fig. 4 refers to the linear period decrease. The residuals after the upward parabolic change was removed are displayed in the lower panel of Fig. 4.

## 3. Photometric solutions with the W-D method

The light curves of V12 displayed in Figs 2 and 3 indicate that it is a total-eclipse binary. The multi-color light curves of the binary are very useful to determine reliable photometric parameters. Therefore, to derive photometric elements and to understand the evolutionary state of the binary star, those light curves were analyzed simultaneously with the W-D program (Wilson & Devinney 1971; Wilson 1979, 1990). The phases of those observations are calculated with the following ephemeris,

$$
Min.I(HJD) = 2456575.24687 + 0^d.3916863 \times E, \qquad (3)
$$

where the initial epoch is one of our times of light minimum, while the orbital period is from Eq. (2). The corresponding phased light curves are shown in Fig. 5.

By considering the B-V=0.636, the temperature for star 1 (star eclipsed at primary light minimum) was fixed as ($T_1 = 5862$) (Cox 2000). Because both components are late-type stars, the gravity-darkening coefficients $g_1 = g_2 = 0.32$ and the bolometric albedo $A_1 = A_2 = 0.5$ were used. The bolometric limb-darkening coefficients $x_{1bolo}$ and $x_{2bolo}$, and the passband-specific limb-darkening coefficients were chosen from Van Hamme (1993) and are listed in Table 3. For a detailed treat of limb darkening, we used the square-root functions for both the bolometric and bandpass limb-darkening laws. During the solution, we found that solutions converged at mode 3, and the adjustable parameters are: the orbital inclination i; the mean temperature of star 2, $T_2$; the monochromatic luminosity of star 1, $L_{1B}$, $L_{1V}$, $L_{1R}$, and $L_{1I}$; and the dimensionless potential ($\Omega_1 = \Omega_2$ for mode 3).

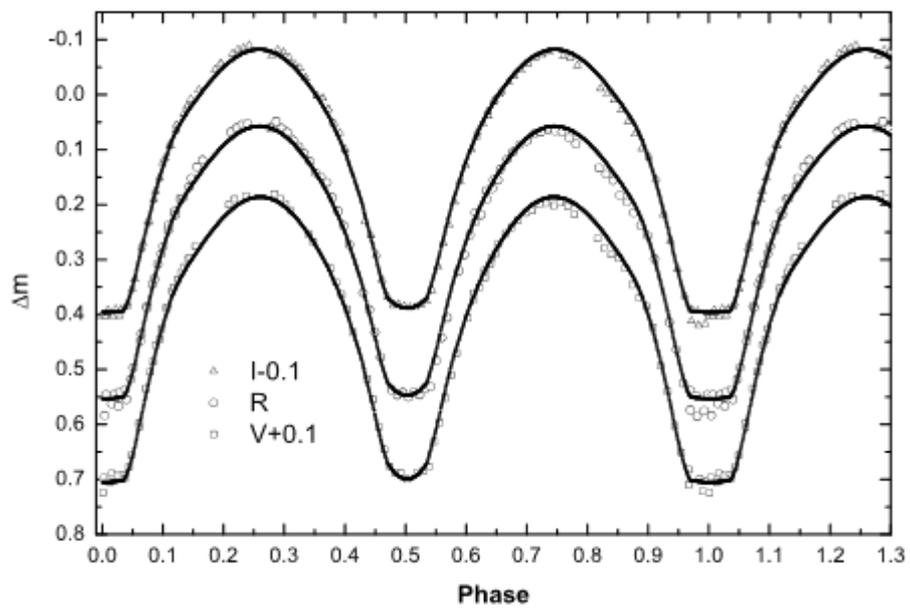

Fig. 5.— Theoretical light curves calculated without considering dark spots. Open squares, circles, and triangles refer to light curve in V, R, and I bands, respectively. It is shown that those theoretical light curves can not fit the observations around two light maxima well.

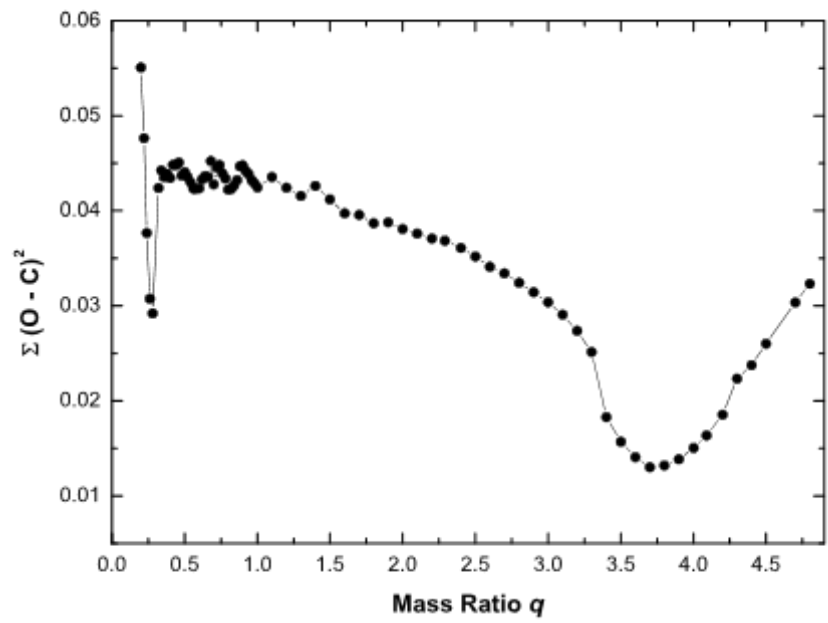

Fig. 6.— The relation between Σ and q.

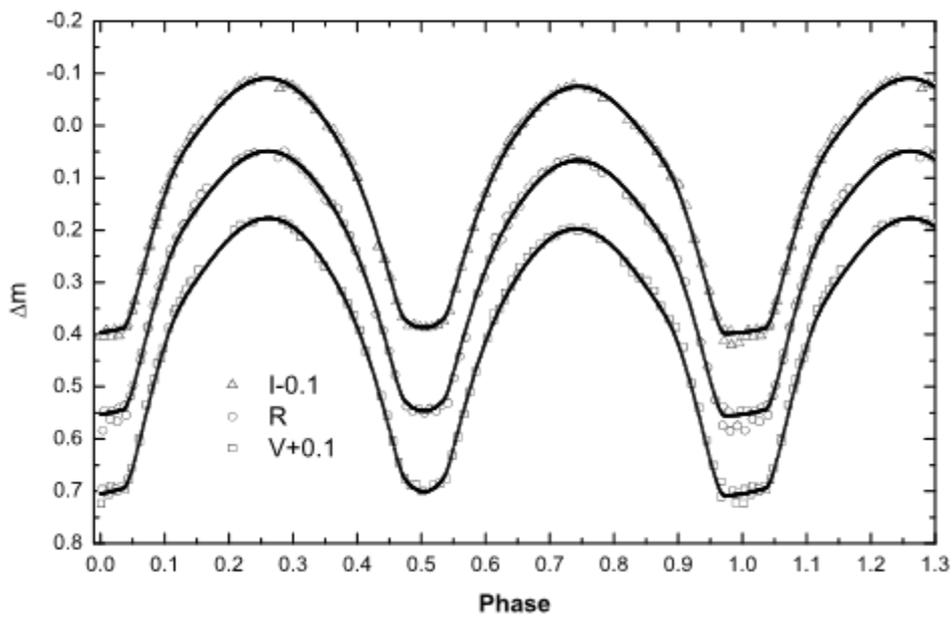

Fig. 7.— Observed (symbols) and theoretical (solid lines) light curves computed with one dark spot on the secondary component that was proposed to explain the positive O'Connell effect. Symbols are the same as those in Fig. 5.

Since no photometric solutions of V12 were obtained before, a q-search method is used to determine the mass ratio firstly. We search for solutions with mass ratio from 0.2 to 4.8, and 75 sets of solution are derived. The relation between the resulting sum Σ of weighted square deviations and q is plotted in Fig. 6. Two minimum values are found at q=0.28 and q=3.7. The solution at q=0.28 indicates that V12 is an A-type contact binary, while the other at q=3.7 reveals it is a W-type system. It is interesting that there is a relation between the two values, i.e., $q_A$ = 0.28 is close to ~$1/q_W$ = 0.27. However, as shown in Fig. 6, the minimum at q=3.7 is much lower than that at q=0.28. Therefore, we chose q=3.7 as the initial value of q and consider it as an adjustable parameter. Then, a differential correction is performed and final solutions were obtained. The photometric solutions are listed in Table 3 and the theoretical light curves are displayed in Fig. 5. Our solution suggests that V12 is a W-type contact binary.

As shown in Fig. 5, the theoretical light curves fitting and the observations are not good around the two maxima. Since both components are fast-rotating solar-type stars, they should show solar-like activity including photospheric dark spots. Therefore, the asymmetry of the light curves can be plausibly explained as the presence of dark spot. In the W-D program, each dark spot has four parameters: the latitude of spot center ($\theta$) in degree, the longitude of spot center ($\varphi$) in degree, spot angular radius ($r$) in radian, and the spot temperature factor ($T_f = T_d/T_0$ where $T_f$ is the ratio between the spot temperature $T_d$ and the photosphere surface temperature $T_0$ of the star). The most widely accepted explanation for the W-type systems was proposed by Mullan (1975). This author introduced cool spots on the more massive components (see also Pribulla et al. 2003). Therefore, a dark spot on the more massive component was proposed to explain the the asymmetry of the light curves, i.e., the positive O'Connell effect. The theoretical light curves are shown in Fig. 7 and the parameters of the dark spot are listed in Table 3. The corresponding geometric structures at different phases are plotted in Fig. 8.

## 4. Discussions and conclusions

Photometric solutions derived in previous section suggest that V12 is a contact binary system with a degree of contact of $f$ = 43.0%. The mass ratio was determined as $1/q$ = 0.26. The color index of B-V=0.636 suggests the system is a G2-type main-sequence star. The mass of the more massive component is estimated as $M_2$ = 1.0$M_\odot$ (Cox, 2000). Then, the mass of the less massive one can be estimated as $M_1$ = 0.26$M_\odot$ by using the value of q. The high orbital inclination of 83.3($\pm$0.3) indicates that the eclipse during the primary minimum is occultation and the derived parameter are reliable. The positive O'Connell effect was explained as the presence of a dark spot on the more massive component. It is found that the temperature of the dark spot is about 1170K lower than the stellar photosphere, while it covers 1.6% of the total photospheric surface.

Table 3: Photometric solutions of V12 in NGC 7789.

| Parameters | Photometric elements Unspotted | uncertainties | Photometric elements Spotted | uncertainties |
|---|---|---|---|---|
| $g_1 = g_2$ | 0.32 | fixed | 0.32 | fixed |
| $A_1 = A_2$ | 0.50 | fixed | 0.50 | fixed |
| $x_{1bolo} = x_{2bolo}$ | 0.185 | fixed | 0.185 | fixed |
| $y_{1bolo} = y_{2bolo}$ | 0.534 | fixed | 0.534 | fixed |
| $x_{1V} = x_{2V}$ | 0.229 | fixed | 0.229 | fixed |
| $y_{1V} = y_{2V}$ | 0.607 | fixed | 0.607 | fixed |
| $x_{1R} = x_{2R}$ | 0.104 | fixed | 0.104 | fixed |
| $y_{1R} = y_{2R}$ | 0.644 | fixed | 0.644 | fixed |
| $x_{1I} = x_{2I}$ | 0.028 | − | 0.028 | fixed |
| $y_{1I} = y_{2I}$ | 0.623 | − | 0.623 | fixed |
| $Phaseshift$ | 0.0026 | ±0.0002 | 0.0039 | ±0.0002 |
| $T_1$ | 5862K | fixed | 5862K | fixed |
| $T_2$ | 5653K | ±10K | 5686K | ±7K |
| $q$ | 3.766 | ±0.012 | 3.848 | ±0.009 |
| $i$ (°) | 83.5 | ±0.5 | 83.6 | ±0.4 |
| $\Omega_1 = \Omega_2$ | 7.4017 | ±0.0095 | 7.4417 | ±0.0095 |
| $\Omega_{in}$ | 7.6134 | − | 7.7174 | − |
| $\Omega_{out}$ | 6.9723 | − | 7.0755 | − |
| $L_1/(L_1 + L_2)(V)$ | 0.2681 | ±0.0014 | 0.2622 | ±0.0010 |
| $L_1/(L_1 + L_2)(R)$ | 0.2620 | ±0.0011 | 0.2572 | ±0.0009 |
| $L_1/(L_1 + L_2)(I)$ | 0.2576 | ±0.0009 | 0.2536 | ±0.0007 |
| $r_1(pole)$ | 0.2658 | ±0.0013 | 0.2684 | ±0.0011 |
| $r_1(side)$ | 0.2787 | ±0.0016 | 0.2820 | ±0.0014 |
| $r_1(back)$ | 0.3240 | ±0.0033 | 0.3326 | ±0.0031 |
| $r_2(pole)$ | 0.4778 | ±0.0010 | 0.4831 | ±0.0009 |
| $r_2(side)$ | 0.5191 | ±0.0014 | 0.5264 | ±0.0012 |
| $r_2(back)$ | 0.5477 | ±0.0018 | 0.5563 | ±0.0016 |
| $f$ | 33.0% | ±2.5% | 43.0% | ±2.2% |
| $\theta$ (°) | − | | 85.94 | ±34.13 |
| $\psi$ (°) | − | | 240.64 | ±8.05 |
| $r$(rad) | − | | 0.1985 | ±0.1550 |
| $T_f$ | − | | 0.8000 | ±0.1797 |
| $\sum (O - C)_i^2$ | 0.01518 | | 0.01504 | |

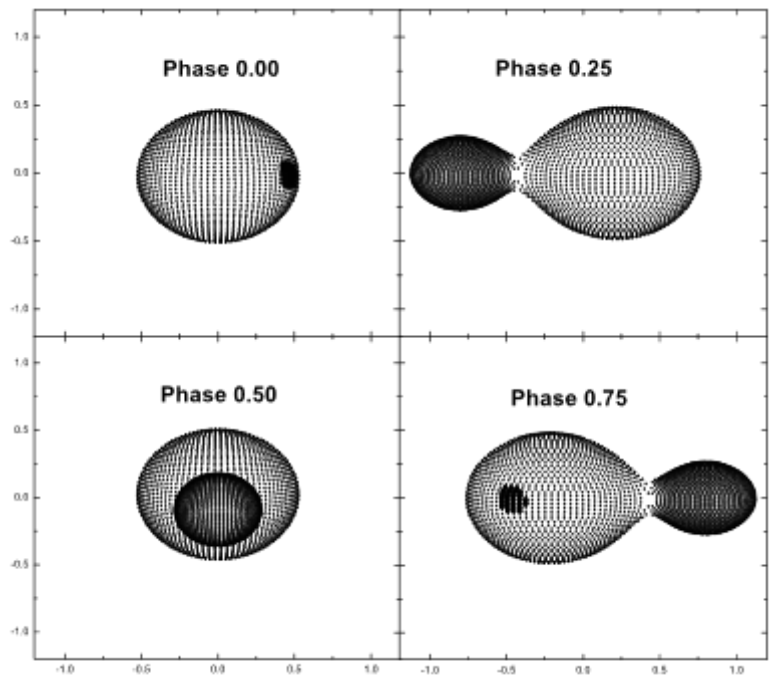

Fig. 8.— Geometrical structure of V12 at phases of 0.0, 0.25, 0.5, and 0.75, respectively. One dark spot on the secondary (the more massive component) is also shown in the panels.

The general trend of the $O - C$ curve in Fig. 4 (solid line) shows an upward parabolic variation indicating a long-term period increase at rate of $\dot{P} = +2.48(\pm0.17)\times10^{-6}$ days/year. It can be explained by the mass transfer from the less massive component to the more massive one. By considering a conservative mass transfer, a calculation with the equation,

$$\dot{P}/P = -3\dot{m}(1/M_1 - 1/M_2), \qquad (4)$$

yields a mass transfer rate of $dm/dt = 2.9\times10^{-7} M_-$/year. However, as shown in the upper panel of Fig. 4, the period increase is only supported by several eclipse times. Moreover, since the rate of the long-term period change is very large, the upward parabolic change may be only a part of a long-period cyclic oscillation that may be caused be the presence of a third body. To confirm the conclusion, more times of light minimum are required in the future.

The origin and evolution of W UMa-type contact binaries were discussed by several investigators (e.g., Huang 1966, van't Veer 1979, Vilhu 1982, Guinan & Bradstreet 1988, Eggen & Iben 1989, and Bradstreet & Guinan 1994). These authors considered that this type of binaries evolve into contact configuration from initially detached binaries by angular momentum loss (AML) via magnetic torques from stellar winds. The evolutionary expansion of the more massive component together with the orbital shrinkage due to AML should result in a mass transfer from the more massive component to the less massive one, and finally a contact system is formed. In this way, it is expected that the pre-contact timescale should be long and intermediate or deep contact binary stars should appear in old open clusters.

Friel & Janes (1993) found the age of NGC 7789 to be about 2Gyr,while the age determined by Gim et al. (1998) is about 1.6 Gyr. Recently, Wu et al. (2007) presented new BATC 13 band photometric results and derived a set of best fitting fundamental parameters for this cluster: an age of $t = 1.4(\pm0.1)$Gyr. On the other hand, the mean age of contact binaries is about 1.61Gyr (e.g., Bilir et al. 2005). This suggests that the age of NGC 7789 is close to the mean lifetime of contact binary stars. The detection of an intermediate-contact binary ($f = 43.0\%$) in the intermediate-age open cluster NGC 7789 reveals that it has a very short pre-contact timescale. One explanation is that it has undergone a collision path to faster evolution and thus have a very short initial orbital period. On the other hand, if there is a third body in the system, it should play an important role by removing angular momentum from the central binary (e.g., Qian et al. 2013a,b,c).

This work is supported by Chinese Natural Science Foundation (No.11133007, No. 11325315, and No.11103074). Also supported by STDF (Science and Technological Development Fund, Ministry for Scientific Research, Egypt), project ID 1335. New CCD photometric observations of V12 were obtained with the 1.88-m Kottamia reflector telescope in Egypt.

## REFERENCES

Azzam, Y. A., Ali, G. B., Elnagahy, F., Ismail, H. A., Haroon, A. and Selim, I., 2010, Astrophysics and space Science Proceeding, 175, 187

Bilir, S., Karatas, Y., Demircan, O., Eker, Z., 2005, MNRAS 357, 497

Biro, I. B., Borkovits, T., Csizmadia, S., Hegedus, T., Klagyivik, P., Kiss, Z. T., Kovacs, T., et al., 2006, IBVS No. 5684

Bradstreet, D. H. & Guinan, E. F., 1994, ASP Conference Series 56, 228

Burbidge, E. M. & Sandage, A., 1958, ApJ 128, 174

Cox, A. N., 2000, Allen's astrophysical quantities, 4th ed. Publisher: New York: AIP Press; Springer

Eggen, O. J. & Iben, I., 1989, AJ 97, 431

Friel, E. D. & Janes, K. A., 1993, A&A 267, 75

Gim M., VandenBerg D. A., Stetson P. B., Hesser J. E., Zurek D. R., 1998, PASP, 110, 1318

Guinan, E. F. & Bradstreet, D. H., 1988, in Dupree, A. K., Lago, M. T. V. T., eds, Formation and Evolution of Low Mass Stars, Kluwer, Dordrecht, P.345

Huang, S.-S., 1966, Ann. d'Astrophys. 29, 331

Jahn K., Kaluzny J., Rucinski S. M., 1995, A&A, 295, 101

Kim C., Park N.-K., 1999, ApSS, 268, 287

Kiron, Y. R., Sriram, K., Vivekananda Rao, P., 2011, RAA 11, 1067

Liu, L., Qian, S.-B., Soonthornthum, B., Zhu, L.-Y., He, J.-J., Yuan, J.-Z., 2007, PASJ 59, 607

Liu, L., Qian, S.-B., Zhu, L.-Y., He, J.-J., Liao, W.-P., Li, L.-J., Zhao, E.-G., Wang, J.-J., 2011, MNRAS 415, 3006

Liu, L., Qian, S.-B., Zhu, L.-Y., He, J.-J., Yuan, J.-Z., Dai, Z.-B., Liao, W.-P., Zhang, J., 2008, PASJ 60, 565

McNamara B. J., Solomom S., 1981, A&AS, 43, 337

Mochejska B. J., Kaluzny J., 1999, Acta Astro., 49, 351

Mullan, D. J., 1975, ApJ 198, 563

O'Connell, D. M. K., 1951, Riverview Pub. 2, 85

Pribulla, T., Kreiner, J. M., Tremko, J., 2003, Contributions of the Astronomical Observatory Skalnaté Pleso 33, 38

Qian, S.-B., Liu, L., Kreiner, J. M., 2006a, NewA 12, 117

Qian, S.-B., Liu, L., Soonthornthum, B., Zhu, L.-Y., He, J.-J., 2006b, AJ 131, 3028

Qian, S.-B., Liu, L., Soonthornthum, B., Zhu, L.-Y., He, J.-J., 2007, AJ 134, 1475

Qian, S.-B., Liu, N.-P., Li, K., He, J.-J., Zhu, L.-Y., Zhao, E. G., Wang, J.-J., Li, L.-J., Jiang, L.-Q., 2013c, ApJS 209, 13

Qian, S.-B., Shi, G., Fern´andez Laj´us, E., Di Sisto, R. P., Zhu, L.-Y., Liu, L., Zhao, E.-G., Li, L.-J., 2013a, ApJ 772, L18

Qian, S.-B., Zhang, J., Wang, J.-J., Zhu, L.-Y., Liu, L., Zhao, E. G., Li, L.-J., He, J.-J., 2013b, ApJS 207, 22

Sriram, K., Vivekananda Rao, P., 2010, RAA 10, 159

van Hamme, W., 1993, AJ 106, 2096

van't Veer, F., 1979, A&A 80, 287

Vilhu O., 1982, A&A 109, 17

Wilson, R. E., 1990, ApJ 356, 613

Wilson, R. E., 1994, PASP 106, 921

Wilson, R. E. & Devinney, E. J., 1971, ApJ 166, 605

Wu, Z.-Y., Zhou, X., Ma, J., et al., 2007, AJ 133, 2061

Zhang, X.-B., Deng, L.-C., 3917, & Zhou, X. 2003, ChJAA(Chin. J. Astron. Astrophys.), 3, 151

Zhu, L. Y., Qian, S. B., Soonthornthum, B., Liu, L., He, J. J., Liu, N. P., Zhao, E. G., Zhang, J., Wang, J. J., 2014, AJ (in press)